\documentclass[12pt]{iopart}

\begin{document}

\newcommand{\phib}{\ensuremath{\overline{\phi}}}
\newcommand{\rhob}{\ensuremath{\overline{\rho}}}
\newcommand{\bq}{\ensuremath{{\bf q}}}
\renewcommand{\cal}{\ensuremath{\mathcal}}
\newcommand{\bqp}{\ensuremath{{\bf q'}}}
\newcommand{\bbq}{\ensuremath{{\bf Q}}}
\newcommand{\bp}{\ensuremath{{\bf p}}}
\newcommand{\brn}{\ensuremath{{\bf r}}}
\newcommand{\bpp}{\ensuremath{{\bf p'}}}
\newcommand{\bk}{\ensuremath{{\bf k}}}
\newcommand{\bx}{\ensuremath{{\bf x}}}
\newcommand{\bxp}{\ensuremath{{\bf x'}}}
\newcommand{\by}{\ensuremath{{\bf y}}}
\newcommand{\byp}{\ensuremath{{\bf y'}}}
\newcommand{\bxpp}{\ensuremath{{\bf x''}}}
\renewcommand{\rmd}{\ensuremath{d}}
\newcommand{\intk}{\ensuremath{{\int \frac{d^3\bk}{(2\pi)^3}}}}
\newcommand{\intq}{\ensuremath{{\int \frac{d^3\bq}{(2\pi)^3}}}}
\newcommand{\intqp}{\ensuremath{{\int \frac{d^3\bqp}{(2\pi)^3}}}}
\newcommand{\intp}{\ensuremath{{\int \frac{d^3\bp}{(2\pi)^3}}}}
\newcommand{\intpp}{\ensuremath{{\int \frac{d^3\bpp}{(2\pi)^3}}}}
\newcommand{\intx}{\ensuremath{{\int d^3\bx}}}
\newcommand{\intxp}{\ensuremath{{\int d^3\bx'}}}
\newcommand{\intxpp}{\ensuremath{{\int d^3\bx''}}}
\newcommand{\drho}{\ensuremath{{\delta\rho}}}
\newcommand{\rhoh}{\ensuremath{{\hat{\rho}}}}
\newcommand{\rhot}{\ensuremath{{\tilde{\rho}}}}
\newcommand{\fh}{\ensuremath{{\hat{f}}}}
\newcommand{\phih}{\ensuremath{{\hat{\phi}}}}
\newcommand{\thetah}{\ensuremath{{\hat{\theta}}}}
\newcommand{\etah}{\ensuremath{{\hat{\eta}}}}
\newcommand{\nh}{\ensuremath{{\hat{n}}}}
\newcommand{\0}{\ensuremath{{(\bk,\omega)}}}
\newcommand{\x}{\ensuremath{{(\bx,t)}}}
\newcommand{\xp}{\ensuremath{{(\bx',t)}}}
\newcommand{\xtp}{\ensuremath{{(\bx',t')}}}
\newcommand{\xtpp}{\ensuremath{{(\bx'',t')}}}
\newcommand{\xttpp}{\ensuremath{{(\bx'',t'')}}}
\newcommand{\xtpn}{\ensuremath{{(\bx',-t')}}}
\newcommand{\xtppn}{\ensuremath{{(\bx'',-t')}}}
\newcommand{\xn}{\ensuremath{{(\bx,-t)}}}
\newcommand{\xpn}{\ensuremath{{(\bx',-t)}}}
\newcommand{\xppn}{\ensuremath{{(\bx',-t)}}}
\newcommand{\xpp}{\ensuremath{{(\bx'',t)}}}
\newcommand{\xxp}{\ensuremath{{(\bx,t;\bx',t')}}}
\newcommand{\Crr}{\ensuremath{{C_{\rho\rho}}}}
\newcommand{\opn}{\ensuremath{\rho}}
\newcommand{\Crf}{\ensuremath{{C_{\rho f}}}}
\newcommand{\Crt}{\ensuremath{{C_{\rho\theta}}}}
\newcommand{\Cff}{\ensuremath{{C_{ff}}}}
\newcommand{\Cffh}{\ensuremath{{C_{f\fh}}}}
\newcommand{\Ct}{\ensuremath{{\dot{C}}}}
\newcommand{\Ctt}{\ensuremath{{\ddot{C}}}}
\newcommand{\Crrp}{\ensuremath{{\dot{C}_{\rho\rho}}}}
\newcommand{\Crfp}{\ensuremath{{\dot{C}_{\rho f}}}}
\newcommand{\Crtp}{\ensuremath{{\dot{C}_{\rho\theta}}}}
\newcommand{\Cffp}{\ensuremath{{\dot{C}_{ff}}}}
\newcommand{\Crrpp}{\ensuremath{{\ddot{C}_{\rho\rho}}}}
\newcommand{\thetab}{\ensuremath{{\overline{\theta}}}}
\newcommand \be  {\begin{equation}}
\newcommand \bea {\begin{eqnarray} \nonumber }
\newcommand \ee  {\end{equation}}
\newcommand \eea {\end{eqnarray}}

\title{Dynamics of interacting
particle systems: stochastic process and field theory}
\author{Alexandre Lef\`evre and Giulio Biroli}
\address{
Service de Physique Th{\'e}orique,
Orme des Merisiers -- CEA Saclay, 91191 Gif sur Yvette Cedex, France.}

\begin{abstract}
We present an approach to the dynamics of interacting particle
systems, which allows to derive path integral formulas from purely
stochastic considerations. We show that the resulting field theory is
a dual version of the standard theory of Doi and Peliti. This clarify both
the origin of the Cole-Hopf map between the two approaches
and the occurence of imaginary noises in effective Langevin equations
for reaction-diffusion systems.
The advantage of our approach is that it focuses directly on the density
field. We show some applications, in particular on the Zero Range Process,
hydrodynamic limits and large deviation functional.
\end{abstract}

\maketitle
\section{Introduction}
\subsection{Generalities}
Many problems of current interest in statistical physics of out of equilibrium systems involve
strongly interacting particles
exhibiting non trivial collective phenomena. One example is that of
supercooled liquids, where the dynamics
slows down dramatically as the glass transition is approached, due to the
increasingly collective nature
of the dynamics, see e.g.\cite{tar}. Another example is given by systems of diffusing
particles that branch and/or annihilate;
depending on the relative strength of these effects a variety of
non-equilibrium transitions and anomalous scaling
behaviour appear~\cite{hin}. A third example is provided by
systems driven out of equilibrium by external sources.
A celebrated example is the one dimensional asymmetric exclusion process
for which one finds phase transitions between different
non equilibrium steady states.

Developing theoretical techniques for such
difficult problems is of great importance, in view of the
diversity of situations in which they appear.

A natural framework to study these collective phenomena is field
theory which has been crucial in understanding equilibrium phase
transitions. In the context of non-equilibrium systems, it has
been already applied successfully to reaction-diffusion systems.
It have been also crucial to get an handle of strong or
intermediate coupling problems where no perturbative technique is
at disposal. Two examples are the application of the Exact
Renormalisation Group to the pair contact process \cite{car,del}
and the Mode Coupling Theory of the glass transition \cite{and}.

A field theoretical formulation of interacting particle systems which has become
standard is based on
the Do\"i-Peliti formalism (DP))~\cite{doi,pel}. Starting from a second quantization
representation of the Master equation, one obtains,
after a rather elaborate coherent state representation, a field theory
representation in terms of two fields $\phi$
and $\phih$ (see below). This has furnished the starting point of a very
large number of studies \cite{car}, including Exact Renormalisation Group
calculations. However, besides its intrinsic difficulty, the formalism
is not transparently related to stochastic equations for the particle evolutions.
Actually, the action of the field theory corresponds to a reasonable
looking Langevin equation for the density of
particles, except that the noise is often complex or even pure
imaginary! This suggests that the field $\phi$, despite
its superficial resemblance with the density, in fact lacks a direct
physical interpretation\cite{tau}. Other difficulties arise when one wants to
treat systems of hard core particles or of particles with non-trivial
diffusion constants.The relationship with stochastic equations on the particle
trajectories is particularly important to study the hydrodynamic limit and
make a connection with the large deviation functional techniques developed in~\cite{der,bod,ber}.
The aim of this article is to discuss in full detail
how stochastic questions on particle trajectories are related to field theory.
We will unveil a dual version of the DP field theory that is naturally related to stochastic equations.
This will shed new light on the underlying structure of DP field theory and we allow us
to re-obtain some recent results from a different perspective, e.g.
the large deviation functional of Bertini et al \cite{ber}
We will also present some new applications, e.g. we will derive
the stochastic equations characterizing the dynamics of the
Zero Range Process. 

\subsection{Issues and questions through a simple example}
Let us illustrate what are the main questions and issues we want to address
focusing on a simple example: particles A
diffusing on a lattice with diffusion constant $\gamma$ and
coalescing ($A+A\rightarrow A$) when they meet
on the same site with rate $\lambda$ per unit of time.
Following the DP formalism, which
will be detailed below, the average density of particles in the
systems is given by the average of a field $\phi$ in a path integral
calculation with action
\begin{equation}\label{eqn:2A}
S=\int_{t_i}^{t_f} \rmd t\int
d^Dx\,\left[\phih\left(-\partial_t\phi+\gamma\Delta\phi\right)
+\lambda\phi^2(1-\phih^2)\right]+{\rm\ boundary\ terms},
\end{equation}
after the extra field $\phih$ has been integrated out.
The quadratic part comes from diffusion, while the rest comes from the
pair coalescence. After a shift $\hat \phi \rightarrow \hat \phi +1$,
the above action becomes identical to the one obtained
through the Martin-Siggia-Rose-DeDominicis-Janssen technique from
the Langevin equation:
\be
\partial_t\phi=\gamma\Delta\phi-2\lambda \phi^2+\eta \qquad \langle\eta(x,t)\eta(x',t')\rangle=-2\lambda \phi^2
\ee
The first two terms of the RHS are exactly what one would expect naively:
a diffusion term plus an annihilation term. The problem is that the white noise has a negative
variance implying that $\eta$ is purely imaginary! This problem, first observed in~\cite{jan}
demonstrates that the field $\phi$ is not equal to the density
field, as can be seen from a direct computation.
Furthermore, it suggests that either the physical sources of
fluctuations in the system are not Gaussian or $\phi$ is not a direct
probe of fluctuations, or both. Still, stochastic equations for the density field
certainly exist. Actually, in many cases one starts from a phenomenological
stochastic equations to get the field theory and not the other way round. But then,
understanding the relationship between stochastic equations for the particle
trajectories and the DP field theory is crucial. This is the main aim of our article which is based on our joint recent work \cite{and} with A. Andreanov and
J.-P. Bouchaud.

The main questions we want to address are:
\begin{itemize}
\item What are the stochastic equations governing the evolution of particles?
\item How are they related to field theory? In particular, starting from these
stochastic equations and using the MSRDJ technique what type of field
theory is obtained?
How is this field theory related to the standard DP one?
\item What is this field theory useful for?
\end{itemize}

In the following we will answer the first two questions in detail. As for the third,
we will show some applications, discuss some other promising ones and hope that
the readers will find new ones.


\subsection{Examples of systems of interacting particles}
In the following we will introduce different classes of systems to which
our theoretical approach applies and on which we shall focus on
 in the next sections.
\subsubsection{Diffusion on a lattice and the Zero Range Process}
The simplest example is that of non (or very weakly) interacting
particles diffusing on the hypercubic lattice $\Lambda^D$. At each
infinitesimal time step $(t,t+dt)$
a particle jumps to a nearest neighbor with
probability $\gamma \rmd t/z$ where
$z$ is the site connectivity and $\gamma$
the diffusion coefficient which may
depend on the position.

The situation gets considerably more complex when $\gamma$ is made
explicitly dependent on the number of particles at each site or in
the neighbourhood. Such is the case in the Zero Range Process
(ZRP), where a particle at site $i$ jump to any neighbour with
probability $\gamma u(n_i)$, where $n_i$ is occupation number of
site $i$ and $u(n)$ a function which vanishes for $n=0$. This
simple dynamic rule, although leading to factorized steady states,
leads to interesting phenomena such as Bose-Einstein
condensation~\cite{evaa}. Many variants exist, with several
species~\cite{evab}, or diffusivity depending only on the target
site~\cite{god}.

\subsubsection{Interacting particles}
Another class of interesting models is that of point particles interacting
via some potential, which for simplicity we shall consider to be
pairwise. This includes lattice systems where sites cannot be occupied
simultaneously by several particles, provided this constraint is
respected by the initial condition. It also includes systems like
gases, simple liquids or crystals, when molecules can
be approximately considered as pointlike. The Hamiltonian of the
system reads:
\begin{equation}
{\cal H}=\sum_{i<j}v(x_i-x_j).
\end{equation}
There are several possible ways of modelling equilibrium dynamics for
such systems, which are equivalent in the continuum limit. We will focus in
particular on the following two:
\begin{enumerate}
\item Particles jumping on lattice from sites to sites,
 using the Metropolis rule for jump acceptance,
\item Particles in free space obeying the Langevin equation:
\begin{equation}
\frac{dx_i}{\rmd t}=-\sum_{j\neq i} v(x_i-x_j)+\eta_i,
\end{equation}
\end{enumerate}
where $\eta_i$ is a Gaussian white noise with variance $\langle \eta_i(t)
\eta_j(t')\rangle=\delta_{i,j}\delta(t-t')$.

If one chooses $v(x)=0$ for $x>2a$ and $v(x)=\infty$ for $x\leq 2a$ and
choose initial conditions such that all particles are at least at
distances $2a$, the system is equivalent to hard spheres of radius
$a$. Furthermore, if the particles evolve on a hypercubic lattice with
spacing $2a$, this is equivalent to condition of single site
occupation. In the so called Exclusion Process,
Symmetric (SEP) or ASymmetric (ASEP), particles diffuse freely
on a lattice with the restriction that they cannot overlap and (in
higher than one dimension) must avoid each other in order to
cross. The system exchanges particles with some infinite reservoirs at
its boundaries, with some rates, or may be closed. As shown for
instance in \cite{der} for the one dimensional case,
the role of boundary conditions is crucial.

In these models, as in those described in the previous paragraph, the
number of particles is conserved.
In a continuous description, this is expressed by
a continuity equation $\partial_t \rho=-\nabla\cdot J$, where $J$ is the
local particle current.

\subsubsection{Reaction-diffusion systems}
In many situations, where particles may
appear, disappear, or  be changed into something else, the number of
particles is not conserved and no continuity equation can be written,
making hydrodynamic descriptions in principle more complicated.
Paradigmatic models for such phenomena are given by particle
assemblies with reaction and diffusion.
For instance molecules deposed on a substrate, in a gas,
or in a porous material,
may diffuse and react chemically. The problem is in general to study
the concentration of molecules in the non equilibrium steady state.
Similar systems are those involving natural species, with birth and
death processes, as well as predation, or
epidemic spreading, where contamination occurs at contact.
In general, such systems can be easily mapped onto systems with
reaction-diffusion. If we denote by $A$ and $B$ two different species,
several reaction processes are possible:
\begin{enumerate}
\item birth: $\emptyset \rightarrow A$,
\item death: $A\rightarrow\emptyset$,
\item coalescence: $2A\rightarrow A$, or more generally
  $mA\rightarrow nA$, with $m>n$,
\item contamination: $A+B\rightarrow 2A$, or $pA+qB\rightarrow
  p'A+q'B$, with $p+q=p'+q'$ and $q'<q$,
\item transmutation: $A\rightarrow B$
\item death at contact: $A+B\rightarrow A$,
\end{enumerate}
the list being not exhaustive. Generically, when they contribute,
these processes occur at some rate $\lambda$ per particle per unit of time.

A common simplification is to allow multiple occupation of sites
and to consider only local reactions occurring at the same site.
One can in principle consider particles with hard core repulsion
or interactions occurring when the particles occupy neighbouring
sites or compact clusters.


\section{Stochastic process and field theory
for interacting particle systems}

In the following we show how to derive the field theory
{\it directly} from the very definition of the stochastic process.
This will bypass completely all the technical machinery based on coherent
states used for DP field theory and it will make clear the relation between
the stochastic process and field theory.
\subsection{Basic processes}
\subsubsection{Particle disintegration}
Let us start with the simple situation of a single site, occupied
at initial time $t_0=0$ by $N_0$ particles. The time is cut in
intervals $[t_i,t_{i+1}]$ ($i=0,\cdots, N-1$) of length $\rmd t$,
during which each particle disintegrates with probability
$\lambda\, \rmd t$. Our goal is to characterize the dynamic
distribution of the number of particles $n(t)$ during the whole
time interval $[0,T]$. To this aim, we introduce an auxiliary
``jump process'' $J(t_i)=n(t_{i+1})-n(t_i)$ and a conjugated field
$\nh(t_i)$, which will probe the generating function
$Z[\{\nh(t_i)\}]=\langle e^{\rmd t \sum_i \nh(t_i)J(t_i)}\rangle$.
For a reason which will be clear below, we temporarily forget that
$J$ is related to the variations of the number of particles. Its
statistics is very simple: $J(t_i)=-1$ with probability
$n(t_i)\lambda\, \rmd t$ and $J(t_i)=0$ with probability
$1-n(t_i)\lambda\, \rmd t$. This gives
\begin{eqnarray}
Z[\{\nh(t_i)\}]&=\prod_i \left(1-n(t_i)\lambda\,\rmd t
+n(t_i)\lambda\,\rmd t\,e^{-\nh(t_i)}\right)\\
&\approx e^{\lambda\,\rmd t \sum_i n(t_i)\left[e^{-\nh(t_i)}-1\right]},\nonumber
\end{eqnarray}
when $N\rightarrow\infty$ and $T=N\,\rmd t$ is kept fixed.
Following the standard approach developed by Martin, Siggia, Rose,
Janssen and de Dominicis (MSRJD) \cite{msr}, the average of any
observable ${\cal
  O}[\{n(t_i)\}]$, {\it with fixed initial conditions} is given by the
following formula:
\begin{equation}\label{eqn:av}
\langle {\cal O}\rangle=\frac{1}{Z}\langle\int
\{\rmd n(t_i)\}\,{\cal
  O}[\{n(t_i)\}]\,\prod_i\delta\left(n(t_{i+1})-n(t_i)-J(t_i)\right)\rangle_J
\end{equation}
where $Z$ is such that $\langle 1\rangle=1$. Here, the number of
particles is ensured to be an integer, as well as its variations,
thanks to the delta functions and to the initial conditions.
Remark that no infinitesimals like $\rmd t$ appear explicitly yet,
as they are hidden in the distribution of the $J$'s.

Using imaginary Fourier representations of the Dirac deltas, the
average in (\ref{eqn:av}) becomes
\begin{eqnarray}
&\langle {\cal O}\rangle=
\frac{1}{Z}\langle\int
\{\rmd n(t_i)\rmd\nh(t_i)\}\,{\cal
  O}[\{n(t_i)\}]\, e^{\sum_i\nh(t_i)\left(n(t_{i+1})-n(t_i)-J(t_i)\right)}\rangle
\\ \nonumber
&\frac{1}{Z}\int
\{\rmd n(t_i)\rmd\nh(t_i)\}\,{\cal
  O}[\{n(t_i)\}]\,\langle e^{\sum_i\nh(t_i)\left(n(t_{i+1})-n(t_i)+\lambda
 \,\rmd t\sum_i n(t_i)\left[e^{-\nh(t_i)}-1\right]\right)}\rangle.
\end{eqnarray}
Taking the continuous time limit (i.e. $\rmd t\rightarrow 0$), one gets
that the probability of observing a given path $\{n(t')\}_{t'\in
  [0,T]}$ is
\begin{equation}
\frac{1}{Z}\int\{\rmd\nh(t_i)\}\,e^{-S[\{n,\nh\}]},
\end{equation}
with
\begin{equation}\label{eqn:des}
S[\{n,\nh\}]=-\int_0^T \rmd t\,\left(
-\nh(t)\partial_t n(t)+\lambda\,n(t)\left[e^{-\nh(t)}-1\right]
\right).
\end{equation}
The action (\ref{eqn:des}) is made of two parts. The first one, with
time derivative, is a ``kinetic'' term, which r\^ole is to fix the now
infinitesimal increments of the field $n$. The second one comes
from the jump process $J$ and encodes the dynamic fluctuations. This
structure in two parts is generic, and thus the steps made above,
which consist in obtaining the probability distribution of the paths
from the generating function $Z[\{\nh(t_i)\}]$ will be skipped in the
other examples. Remark here that the introduction of the field
$\nh$ allows to use real values for $n$ instead of integers only.

The generating function generates cumulants of the distributions of
the variation of particle numbers. In particular, the formula
\begin{equation}\label{eqn:gen}
  \langle \partial_t n(t)\rangle=\left.\frac{\delta Z[\{\nh\}]}{\delta\nh(t)}\right|_{\nh(t)=0}
\end{equation}
will be useful for the derivation of hydrodynamic equations.

\subsubsection{Particle creation}
Another simple process has been described briefly in the
introduction. The calculation follows the lines of the previous
paragraph, and gives
\begin{equation}\label{eqn:cre}
S[\{n,\nh\}]=\int_0^T \rmd t\,\left(
\nh(t)\partial_t n(t)+\lambda\,\left[e^{\nh(t)}-1\right]
\right).
\end{equation}

\subsubsection{Diffusion}
The diffusion process
can be either seen as the motion of a particle performing a sequence
of jumps on 
connected sites, or as an exchange of particles between nearest
neighbours. As we are interested in the density
fluctuations, we adopt the second point of view, which amounts at
considering only two neighbouring sites $1$ and $2$, with initial
occupations $n_1(0)$ and $n_2(0)$. Particles may hop back and forth
with rate $\gamma\rmd t$. The (integer) numbers of
particles on the two sites at time $t$ are $n_1(t)$ and $n_2(t)$.
The variation of $n_k$ between $t$ and $t+\rmd t$ will again be noted
$dJ_k(t)$, while particles hop from $1$ to $2$ with rate $W_{12}$
and from $2$ to $1$ with rate $W_{21}$. Of course,
$dJ_1(t)$ and $dJ_2(t)$ are strongly correlated since a particle
leaving site $1$ lands on site $2$ and vice-versa.
More precisely, $dJ_1(t)=-dJ_2(t)=+1$ with probability $n_2(t) W_{21} \rmd t$,
$dJ_1(t)=-dJ_2(t)=-1$ with probability $n_1(t) W_{12} \rmd t$,
and $dJ_1(t)=dJ_2(t)=0$ otherwise. As before, the dynamical action is
obtained through use of the generating function, leading to
\be
\fl S[\{n,\hat n\}] = -\int\rmd t \left\{-\hat n_1 \partial_t
n_1-\hat n_2 \partial_t n_2+ n_1 W_{12}
(e^{\hat n_2 - \hat n_1} -1) +  n_2 W_{21} (e^{\hat n_1 - \hat n_2} -1)
\right\}.
\ee
On a lattice, the total MSRJD action reads:
\begin{equation}\label{eqn:dif}
S[\{n,\hat n\}] = -\int \rmd t \left\{-\sum_i \hat n_i \partial_t
n_i+ \sum_{\langle ij\rangle} n_i W_{ij}
(e^{\hat{n}_j - \hat n_i} -1) \right\},
\end{equation}
where the brackets restrict the summation to pairs of nearest
neighbours. Later, we shall use this action to get continuum limits.

\subsection{Interacting particles and Dean equations}
Let us consider particles diffusion on a lattice with fixed hopping
rate, but feeling a potential $U_i(t)$ at site $i$. This potential can
be external and time dependent or can also be due to the
interaction of particles at site $i$ with all others, in particular
$U_i(t)=a^D\sum_{j\neq i} v(a|i-j|)n_i(t) n_j(t)$. A natural
stochastic dynamics is to accept hops to neighbouring sites with
the heat bath rule, that is by choosing
\begin{equation}
W_{ij}=\frac{\gamma}{1+e^{-\beta (U_i-U_j)}},
\end{equation}
where $\beta=1/\gamma$ is the inverse temperature. In the continuum limit
$W_{ij}$ can be expanded in powers of $a(i-j)$ and Langevin dynamics
is recovered.

Before doing so, it is important to define the continuum limit
properly. We start with off-lattice particles in a box of very
large size $L$ with given boundary conditions. We then devide this
box into many tiny boxes of size $a$ at position $x=ai$. We use
the number $n_i$ of particles in this box to define the density
$\rho_x=n_i/a^D$ and its conjugate $\rhoh_x=\nh_i$. We choose $a$
very small in order to expand in powers of $a$. At the lowest
useful orders:
\begin{eqnarray}\label{eqn:exp}
e^{\nh_j-\nh_i}-1&=&e^{a
  e_{ij}\cdot\nabla\nh_i+\frac{a^2}{2}(e_{ij}\cdot\nabla)^2\nh_i}-1+o(a^2)\\
&=&a e_{ij}\cdot\nabla\nh_i+\frac{a^2}{2}\left[
(e_{ij}\cdot\nabla)^2\nh_i+(e_{ij}\cdot\nabla\nh_i)^2
\right]+o(a^2),
\end{eqnarray}
where $e_{ij}$ is the unit vector pointing from $i$ to $j$. One can
also expand the expression in $W_{ij}$ and get
\begin{equation}
W_{ij}=\gamma-\frac{1}{2}a e_{ij}\cdot\nabla U_i+o(a).
\end{equation}
In addition, using $x=ia$, one can replace $a^D\sum_i$ by $\int\rmd^D
x$. Terms of order $a$ vanish by symmetry, and thus keeping terms of
order $a^2$ and rescaling $t$ by $a^2$, one gets the following action:
\begin{equation}
\fl
S[\{n,\hat n\}] =-\int\rmd t\int\rmd^D x\left\{
\rhoh_x\left(-\partial_t\rho_x+\gamma
\Delta\rho_x+\nabla\cdot\left[\rho_x\nabla
  U_x\right]\right)+\gamma\rho_x\left(
\nabla\rhoh_x\right)^2
\right\}
\end{equation}
This action can also be obtained from a MSRJD treatment of the
following Langevin equation:
\begin{equation}
\partial_t\rho_x=-\nabla\cdot J_x,
\end{equation}
with a fluctuating current
\begin{equation}\label{eqn:dea}
J_x=-\gamma\nabla\rho_x-\rho_x\nabla
  U_x+\sqrt{\rho_x}\,\xi_x,
\end{equation}
where $\xi_x(t)$ is a Gaussian field with variance
\begin{equation*}
\langle \xi_x(t)\xi_{x'}(t')\rangle=2\gamma
\delta(x-x')\delta(t-t').
\end{equation*}
The 
deterministic part of the current can be expressed in the form
$-\rho_x\nabla\frac{\delta {\cal
    F}}{\delta\rho_x}$, with a free-energy functional
\begin{equation}
{\cal F}[\{\rho\}]=\int\rmd^Dx \left(T\rho_x\ln(a^D\rho_x)+\rho_x
\nabla U_x\right).
\end{equation}
This stochastic differential equation (SDE) for the density field
was first derived by
Dean~\cite{dean} directly from the Langevin equation
for interacting
particles by using It\^o formula.
The free energy ${\cal F}$ is the same as that from
mean-field theory. Our derivation gives further evidence that no
coarse-graining is needed in order to get such free energy. In fact
the interaction $U$ used here is often obtained as an effective
interaction (eg. for colloidal systems) and thus the free energy
functional is actually different from the mean-field one, which
contains the microscopic potential instead.

\subsection{Newtonian dynamics}
Now let us consider systems of particles with Newtonian dynamics and
interacting via a pairwise potential $v$. The equations of motion read
\begin{eqnarray}
\partial_t x_i&=&\frac{p_i}{m}\\
\partial_t p_i&=&-\sum_{j\neq i}\nabla v(x_i-x_j)
\end{eqnarray}
We introduce the density field in phase space:
\begin{equation}
\rho(x,p,t)=\sum_i\delta(x_i(t)-x)\delta(p_i(t)-p).
\end{equation}
It is easy to show that it obeys the deterministic equation
\be
\fl\partial_t\rho(x,p,t)=-\frac{p}{m}\partial_x\rho(x,p,t)
+\int \rmd^D x'\rmd^D p'\, \rho(x',p',t)\nabla V(x-x')\cdot\partial_p\rho(x,p,t),
\ee
all randomness being in the initial conditions. One can write path
integral formulas as before, introducing $\rhoh(x,p,t)$ fields. If one
applies the reverse map of (\ref{eqn:map}) to these fields, one gets
the formulas originally obtained by Doi~\cite{doi}.

\subsection{Exclusion processes}
Dealing with hard cores is in general difficult. A first attempt was
done in~\cite{fvw} Particles with
hard core repulsion may be modelled in two ways. The first one is
to add a pairwise interaction which is infinite if the distance
between the particles is less than their radius and vanishes else.
The second one is to include a constraint in the hopping rate of
the particles on the lattice. From the stochastic approach which
has been developed here, the later is easier to deal with. We thus
start with particles on a lattice with initial occupation number
of site $i$ $n_i(0)\in\{0;1\}$. The
exclusion constraint is propagated by the dynamics, if the hopping
rate from $i$ to $j$ is of the form
\begin{equation}\label{eqn:exc}
n_i W_{ij}=A_{ij}n_i (1-n_j),
\end{equation}
which vanishes if site $i$ is empty or site $j$ is occupied. For
sake of clarity, we choose $A_{ij}=\gamma$, which corresponds to
symmetric exclusion processes. Inserting (\ref{eqn:exc}) into
(\ref{eqn:dif}), expanding like in (\ref{eqn:exp}) and using
$n_j=n_i+a e_{ij}\cdot\nabla n_i+o(a)$, the action describing the
dynamics becomes in the limit $a\rightarrow 0$: \be\label{eqn:sep}
\fl S[\{n,\nh\}]=-a^{-D}\int\rmd t\int\rmd^D x\left\{
-\nh_x\partial_t n_x+\gamma \nh_x\Delta n_x+\gamma
n_x(1-n_x)\left(\nabla\nh_i\right)^2 \right\}, \ee
after rescaling of the microscopic time $t\rightarrow t/a^2$. 
This looks like
the hydrodynamic equation obtained rigourously by Bertini et
al.~\cite{ber} for symmetric exclusion processes - althought the time
rescaling is different. However, our
approach suggests that this equation is valid beyond the
hydrodynamic limit. In addition, although we started from
$n_x\in\{0;1\}$, the measure gives also non-zero weights to paths
such that $n_x(1-n_x)\neq 0$. As we shall see, this is related to
the fact that the hydrodynamic limit corresponds to a saddle point
of the dynamic action which is located on continuous paths, and
that the dynamics on slightly smaller scales is given by
fluctuations around this saddle point.

\subsection{Initial conditions}
Up to now, we have not specified the initial conditions and
considered only the dynamical changes of the density. In many
cases, initial conditions are forgotten after some transient time
and a stationary state is obtained. However, in the case of
exclusion processes, our stochastic approach to the dynamics makes
it mandatory - at least for consistence - to specify initial
conditions with the good occupation numbers, in order to insure
that the paths supporting the measure are consistent with the
dynamic rules. We assume a factorized initial state with $q_i$
particles on site $i$, drawn with a distribution $P_i(q_i)$. It is
useful to define the generating function $g_i(z)=\ln\left(\sum_q
P_i(q)z^q\right)$. In order to specify the initial conditions, we
add a term
$\left\langle\prod_i\delta\left(q_i-n_i(0)\right)\right\rangle$.
Using integral representations of the $\delta$'s, this gives an
extra term to the dynamic action, after averaging:
\begin{equation}
S_{\rm I}[\{n(0),\nh(0)\}]=\sum_i n_i(0)\nh_i(0)-\sum_i
g_i\left(e^{\nh_i(0)}\right).
\end{equation}
A quite generic case is that of Poissonian initial state, with
$P_i(n)=\frac{\rho_{0,i}^n}{n!}e^{-\rho_{0,i}}$, which leads to
\be S_{\rm I}[\{n(0),\nh(0)\}]=\sum_i n_i(0)\nh_i(0)-\sum_i
\rho_{0,i}\left(e^{\nh_i(0)}-1\right). \ee If hard cores prevent
several particles to sit on the same site, one may choose
independent Bernoulli distributions with average $\rho_{0,i}$,
leading to \be S_{\rm I}[\{n(0),\nh(0)\}]=\sum_i
n_i(0)\nh_i(0)-\sum_i \ln\left(
1-\rho_{0,i}+\rho_{0,i}e^{\nh_i(0)}\right). \ee

\subsection{Boundary conditions}
It may happen that changing boundary conditions changes drastically
the dynamical behaviour. One dimensional exclusion processes are
popular examples of such phenomenon. Periodic boundary conditions are
taken into account by imposing space periodicity on the fields. Taking
into account open boundary conditions is more model specific, and we
explain how to do it on the example of an exclusion process in one
dimension. We assume that the system is coupled to two reservoirs at
its ends. Particles are added at boundary points $0$ and
$N$ with rates $\alpha$ and $\gamma$, and removed with rates $\beta$
and $\delta$. As the number of particles is preserved inside the
system, its fluctuations are governed by exchanges of particles with
the reservoirs. This gives an extra term to the dynamical action:
\begin{eqnarray}
\fl &S_{\rm B}[\{n_0,n_N,\nh_0,\nh_N\}]=
-\int\rmd t\,\left\{\alpha
\left(1-n_0(t)\right)\left(e^{\nh_0(t)}-1\right)\right.\\ \nonumber
\fl&+\left.\gamma\left(1-n_N(t)\right)\left(e^{\nh_N(t)}-1\right)
+\beta n_0(t)\left(e^{-\nh_0(t)}-1\right)
+\delta n_N(t)\left(e^{-\nh_N(t)}-1\right)
\right\}.
\end{eqnarray}

\section{Mapping between Doi-Peliti and stochastic approaches}
This section is devoted to show mapping between the above approach and the more
standard one based on the DP method. For self-consistency, we start by
recall briefly the derivation of path integral formula using the
latter method.

\subsection{Doi-Peliti path integral formulation}
Let us consider particles with annihilation
$A+A\rightarrow\emptyset$ on a single site at rate $2\lambda$,
which will provide a benchmark to show our method. The initial
number of particles is $n_0$. The probability of having $n$
particles at time $t$ evolves according to the master equation
\begin{equation}\label{eqn:mas}
\partial_t P(n,t)=\lambda\,
(n+2)(n+1)P(n+2,t)-\lambda\,n(n-1)P(n,t).
\end{equation}
We map this onto a imaginary time Schr\"odinger equation by
introducing a Fock space generated by $n$-particle states
$|n\rangle$ and introducing the state ket of the system at time
$t$:
\begin{equation}
|\psi(t)\rangle=\sum_n P(n,t)|n\rangle.
\end{equation}
We also use the lowering/raising operators $a$ and
$a^+$ according to the standard definition
\begin{eqnarray}
a|0\rangle&=&0\\
a^+|n\rangle&=&|n+1\rangle\\
a|n\rangle&=&n|n-1\rangle,
\end{eqnarray}
which verify the commutation relation $[a,a^+]=1$. The master
equation (\ref{eqn:mas}) now becomes
\begin{equation}
\partial_t|\psi(t)\rangle=-\hat H|\psi(t)\rangle,
\end{equation}
with
\begin{equation}\label{eqn:ham}
\hat H=\lambda\left({a^+}^2-1\right)a^2.
\end{equation}
We obtain the state ket at any time:
\begin{equation}
|\psi(t)\rangle=e^{-\hat H t}|\psi(0)\rangle,
\end{equation}
and
\begin{equation}
P(n,t)=\frac{\langle n|\psi(t)\rangle}{n!}.
\end{equation}
The $n$-particles states verify $\langle
n|m\rangle=\delta_{n,m}n!$. As a consequence one can define coherent
states
\begin{equation}
  |\phi\rangle=\frac{1}{\sqrt{\pi}}e^{-\frac{\phi\phi^*}{2}}e^{\phi
   a^+}|0\rangle
\end{equation}
and
\begin{equation}
  \langle\phi|=\frac{1}{\sqrt{\pi}}e^{-\frac{\phi\phi^*}{2}}\langle 0|e^{\phi^*
   a}.
\end{equation}
The coherent states thus define the (over)complete relation:
\begin{equation}\label{eqn:com}
1=\int\rmd\phi\rmd\phi^* |\phi\rangle\langle\phi|.
\end{equation}
The average of an observable $A(n)$ is given by
\begin{eqnarray}
\langle A\rangle_t&=&\sum_n A(n)P(n,t)\\
&=&\langle P|\hat A|\psi(t)\rangle\\
&=&\langle P|\hat A\, e^{-\hat H t}|\psi(0)\rangle,\label{eqn:obs}
\end{eqnarray}
where
\begin{equation}
\hat A=\sum_n|n\rangle\frac{A(n)}{n!}\langle n|,
\end{equation}
and the ``projection state'' $\langle P|=\langle 0|e^a$ is a left
eigenvector of $a^+$ with eigenvalue $1$. Using the Trotter
formula
\begin{equation}
e^{-\hat H t}=\left(1-\hat H \rmd t+o(\rmd t)\right)^M,
\end{equation}
with $M=t/\rmd t$, and inserting (\ref{eqn:com}) in between all
factors, one obtains a product of terms of the form
\begin{equation}
\langle \phi_{k+1}|\phi_{k}\rangle\left(1-\rmd t\,\frac{\langle\phi_{k+1}|\hat
  H|\phi_{k}\rangle}{\langle \phi_{k+1}|\phi_{k}\rangle} \right).
\end{equation}
It is easy to show that
\begin{equation}
\langle\phi_{k+1}|\hat H|\phi_{k}\rangle=\langle
\phi_k|\phi_{k}\rangle H(\phi^*_{k+1},\phi_{k}),
\end{equation}
where
$H(\phi^*,\phi)$ is obtained by replacing respectively $a^+$ and
$a$ by $\phi^*$ and $\phi$ in the expression of $:\hat H:$ in terms of
$a^+$ and $a$. The expression $:\hat H:$ is obtained from $\hat
H$ by normal ordering, i.e. commuting all operators until all $a$'s
are on the right. In this example (\ref{eqn:ham}) has already been
normal ordered, and thus
\begin{equation}
H(\phi^*,\phi)=\lambda\left({\phi^*}^2-1\right)\phi^2.
\end{equation}
We thus get:
\begin{eqnarray}
\langle A\rangle_t&=\int\prod_{k=0}^M\rmd \phi_k\rmd\phi^*_k\,
\langle P|\hat A|\phi_M\rangle\langle\phi_0|\psi(0)\rangle\\\nonumber
&\times\left\{\prod_{k=1}^M\langle\phi_k|\phi_{k-1}\rangle
\left(1-dt H(\phi^*_k,\phi_{k-1})\right)
\right\}.
\end{eqnarray}
The operator $\hat A$ can always be written in the normal ordered form
$A(a^+,a)=:\hat A(a^+,a):$. In addition, using the identity
$[e^a,f(a^+)]=f(a^++1) e^a$, we get
\be
\langle P|\hat A|\phi_M\rangle=A(1,\phi_M)\,e^{\phi_M}.
\ee
In addition:
\begin{eqnarray}
\langle \phi_0|\psi(0)\rangle&=&\sum_q
P(q)\langle\phi_0|(a^+)^q|0\rangle \\
&=&\sum_q P(q)\phi^{*q}_0\\
&=&e^{g(\phi_0^*)}.
\end{eqnarray}
In the limit $\rmd t\rightarrow 0$, we introduce the continuous time
$s=k\rmd t$ and thus
\be\label{eqn:rd}
\langle A\rangle_t=\int\{\rmd\phi^*\}\{\rmd\phi\}\,
A(1,\phi_t)\,e^{\phi(t)+g(\phi_0^*)+\int_0^t\rmd s\,H(\phi^*(s),\phi(s))
-\int_0^t\rmd s\,\phi(s)\partial_s\phi^*(s)}.
\ee
Diffusion processes can be treated in the same way. The resulting
hamiltonian is
\be\label{eqn:hd}
H_{\rm dif}[\{\phi^*,\phi\}]=\gamma\sum_{\langle
  i,j\rangle}\left(\phi^*_i-\phi^*_j\right)\left(\phi_i-\phi_j\right)
\ee
Correlation functions at different times can be computed in the same
way. For instance:
\be
\langle
n(t_1)\cdots n(t_p)\rangle=\langle P|
a^+ a
\prod_{k=2}^p\left\{ e^{-(t_k-t_{k-1})H(a,a^+)}a^+a \right\}
\sum_q P(q) |q\rangle.
\ee
The average of the $\phi^*$ field is $1$, and in general the shift
$\phib=\phi^*+1$ is carried out, in order to deal with a field with
vanishing average. This also cancels the isolated term $\phi(t)$ in (\ref{eqn:rd}).

\subsection{The Cole-Hopf transformation}
We now show how the action obtained from purely stochastic
considerations can be exactly mapped onto the one derived in the
previous paragraph. The particle number operator is $\hat n=a^+a$.
Thus, it is tempting to express the density field as
$\rho=\phi^*\phi$. However, in order to go from operators $a,a^+$
to fields $\phi,\phi^*$, one must normal order. For instance
$:{\hat n}^2:=(a^+)^2a^2+a^+a$, and thus in the field theoretic
formalism, one should rather choose
$\rho^2=(\phi^*\phi)^2+\phi^*\phi$ instead of
$\rho^2=(\phi^*\phi)^2$. Before dealing with this ambiguity, we
will proceed to a na\"\i ve mapping between fields in the
different formulations, and apply it to situations where it is
correct. Let us define
\begin{eqnarray}\label{eqn:map}
\rho_i(s)&=&\phi^*_i\phi_i(s),\\
\rhoh_i(s)&=&\ln\phi^*_i(s).
\end{eqnarray}
The mapping (\ref{eqn:map}) has unit determinant and thus does not
change the measure. In addition, it is valid both on lattice and
in the continuum. Plugging (\ref{eqn:map}) into (\ref{eqn:hd}),
one gets 
\be 
\gamma\sum_{\langle i,j\rangle} \int_0^t\rmd
s\,\rho_i(s)\left( e^{\rhoh_j(s)-\rhoh_i(s)}-1\right). 
\ee 
This is
precisely the terms resulting from the calculation of the
generating function in the stochastic approach. The extra kinetic
term in (\ref{eqn:rd}) becomes under (\ref{eqn:map}): \be
-\sum_i\int_0^t\rmd s\,\rho_i(s)\partial_s\rhoh_i(s). \ee This
shows that the field defined as $\rho=\phi^*\phi$ is a ``good''
density field, as it formally corresponds to the one of the
stochastic method. The nature of density fluctuations is thus
hidden in the DP formalism. Thanks to the Cole-Hops mapping
detailed above one obtains a dual version of the field theory which
turns out to be the DP one as the reader can easily check for all
processes considered in this paper. There is only one subtle point
needed for this derivation that we shall discuss below. For
concreteness let us come back to the reaction
$A+A\rightarrow\emptyset$. The non-kinetic bulk part of the
dynamic action reads in the DP form: \be\label{eqn:aaDP} S_{\rm
A+A,DP}=-\int_0^t\rmd s\,\lambda\phi^2(s)\left(
1-\phi^*(s)^2\right). \ee The equivalent term in the stochastic
approach is: \be\label{eqn:aaMSRJD} S_{\rm
A+A,MSRJD}=-\int_0^t\rmd s\,\lambda\rho(s)\left(\rho(s)-1\right)
\left(e^{-2\rhoh(s)}-1\right). \ee However, if we use the mapping
(\ref{eqn:map}) into (\ref{eqn:aaDP}), one gets \be\label{65}
-\int_0^t\rmd s\,\lambda\rho(s)^2\left(e^{-2\rhoh(s)}-1\right),
\ee which differs from (\ref{eqn:aaMSRJD}). In order to understand
the origin of this difference, we recall that the fields $\phi$
and $\phi^*$ are quantum fields, as a consequence the products of
fields at the same time is a quantity that depends crucially on
the underlying discretization. For example $\phi^* (t^+) \phi(t^-)
\neq \phi (t^+) \phi^*(t^-)$ as it can be readily checked. In
order to understand which discretization allows one to map the
field theory derived in the previous section to the DP one we
carry a mapping  equivalent to (\ref{eqn:map}) at the operator
level:
\begin{eqnarray}\label{eqn:mapop}
a&=& e^{-\opn^+}\opn\\
a^+&=& e^{\opn^+}.
\end{eqnarray}
Operators $\opn$ and $\opn^+$ have canonical commutation rules:
\begin{eqnarray}
[\opn,\opn^+]&=&a^+\,[a,\ln a^+]\nonumber\\
&=&a^+ \partial \ln a^+/\partial a^+\\\nonumber
&=&1.
\end{eqnarray}
It is thus natural to use them as lowering/raising operators. We
define a new set of Fock space generating vectors as follows:
\begin{eqnarray}
|\tilde{0}\rangle&=&|0\rangle,\\
|\tilde{n}\rangle&=&(\opn^+)^n|\tilde{0}\rangle,
\end{eqnarray}
and corresponding left vectors:
\begin{eqnarray}
\langle \tilde{0}|&=&\langle P|\\
\langle \tilde{m}|&=&\langle\tilde{0}|\rho^m.
\end{eqnarray}
We point out that $\opn$ and $\opn^+$ are not hermitian
conjugates, and thus neither are $\langle\tilde{n}|$ and
$|\tilde{n}\rangle$. We can redo the DP derivation of the path
integral formulas, but using these operators instead of operators
$a$ and $a^+$, writing
\begin{equation}
  \hat H(a^+,a)={\hat H}_{\rm stoc}(\opn^+,\opn).
\end{equation}
Normal ordering with respect to $\opn^+$ and $\opn$ and replacing
respectively $\opn^+$ and $\opn$ with $\rhoh$ and $\rho$, one gets the
equivalent of (\ref{eqn:rd}), where the term at time $s$ is
\begin{equation}\label{eqn:rds}
  -\rhoh(s)\partial_s\rho(s)+H_{\rm stoc}(\rhoh(s),\rho(s)).
\end{equation}
Initial conditions are also taken into account:
\begin{eqnarray}\label{eqn:ins}
\sum_q P(q)\langle\opn^*_0|\tilde{n}\rangle&=&\sum_q P(q)e^{q\opn^*}\nonumber\\
&=&e^{g(e^{\opn^*_0})}.
\end{eqnarray}
Gathering terms from (\ref{eqn:rds}) and (\ref{eqn:ins}), we get
{\em exactly} the dynamic action derived from the purely stochastic
approach. From the point of view of the time discretization the above
derivation clarify what was the origin of the discrepancy between
(\ref{eqn:aaMSRJD}) and (\ref{65}). The field theory derived directly
from the stochastic equation gives, after fields transformation,
$\rho(t)\rho(t)= \lim_{\epsilon \rightarrow 0}
\phi^*(t+4\epsilon) \phi(t+3\epsilon) \phi^*(t+2\epsilon) \phi(t+\epsilon)$.
Instead within the DP field theory the corresponding term is
$\phi^*(t)\phi^*(t) \phi(t)\phi(t) =
\lim_{\epsilon \rightarrow 0}
\phi^*(t+4\epsilon) \phi^*(t+3\epsilon) \phi(t+2\epsilon) \phi(t+\epsilon)
$. As a consequence, in order to map one onto the other one has to
use also the same time discretization.

We remark that we have not tried to justify the existence of
the operators $\opn$ and $\opn^+$. The change (\ref{eqn:mapop}) may
not be well defined at the operator level. However, it is a
posteriori justified by the fact that it gives exactly the action
obtained from another, stochastic, approach. Furthermore, the same result
can be obtained alternatively just by discretizing the field theory.

As a consequence, although path integral formulas can be obtained
both using the simple MSRJD stochastic method or the more abstract
DP one, the previous discussion highlights that care must be taken
in the interpretation of the field theory. Indeed, depending on
the physical situation, the sets of fields $\phi,\phi^*$ and
$\rho,\rhoh$ are more adapted than the other. However, different
sets are adapted to different approaches and going from one to the
other using (\ref{eqn:map}) may be dangerous. In fact this mapping
has been often referred to~\cite{car,jan,fvw,grass,bra}, in a rather
loose way, in particular without attention to the fact that the
commutation rules of the fields are important.

\subsection{Physical origin of fluctuations}
Let us discuss in more detail the origin of the difference between
the two field theories obtained in the previous sections. The
noise generated by the stochastic dynamics is intrinsically
Poissonian at the microscopic level. For example, consider
particles created on a site (empty at $t=0$) with rate $\lambda$
per unit of time. The generating function of $N(t)$, the number of
particle at time $t$
 is $f(s)=\langle e^{s N(t)}\rangle$. It is a straightforward
exercice to get
\begin{equation}
f(s)=\exp\left[\lambda t\left(e^s-1\right)\right]
\end{equation}
in the limit $\rmd t\rightarrow 0$ keeping $t$ fixed. Similar expressions,
are obtained for diffusions and more complicated processes. The intrinsic
Poissonian nature of the noise, i.e. the fact that with very small probability
variables change of a finite amount, makes the logarithm of the generating
function a complicated function of the argument. In particular it leads
generically to exponential terms. This is the main reason
why the action of the
field theory in the $\rho, \hat \rho$ variable is complicated even for simple
processes. On the other hand, the field theory in the $\phi,\phi^*$
variables is simple for simple processes as creation, annihilation or
diffusion. 
In particular for diffusion the action is quadratic. The drawback is that
$\phi$ is not the density field. This is the reason why stochastic equations
on $\phi$ doe not make sense physically and lead to inconsistencies, as
the presence of imaginary noises.

The usefulness of each of the two field theories depends on the
particular physical 
problem one is interested in. One focuses directly on
the {\it physical} variables but it is already complicated (non-Gaussian)
for simple processes. The DP one is instead simple in these cases;
however computing physical observables may be quite complicated because they
correspond to high order multipoint correlation functions. Thus, for
renormalization group treatment of simple processes the DP one seems and
has been proven to be very useful. On the other hand, in cases where
one is interested in two point density correlation functions the other field
theory seems to be more useful because it allows to develop approximations
directly for these functions avoiding to deal with high order multipoint
(more than two points) correlation functions which is in general a
quite cumbersome task. 
Finally, we remark that on sufficiently large length scales the noise
becomes Gaussian.
In this case the field theory on $\rho, \hat \rho$ leads directly to
hydrodynamic descriptions away from critical points.
Hydrodynamic descriptions correspond to coarse-graining the dynamics
on mesoscopic scales much bigger than the correlation length, with
corresponding rescaling of time. From coarse-graining results a law of
large numbers under the form of a large deviation theory, providing
the probability of occurence of coarse-grained trajectories, or of
density profiles in the steady state.

\section{Applications}
We now give some applications of the formalism which we have explained
in details. We start with the zero range process for which we derive
continuum stochastic equations.

\subsection{Zero Range Process}

Zero Range Processes (ZRP) provide a simple model for the study of
particles undergoing an out of equilibrium process.
Indeed, the dynamic rules, although quite
simple may lead to a rather reach phenomenology. In addition, in the
simplest cases, the steady state measure can be computed and is
factorized. For a detailed review, see~\cite{evaa}.

\subsubsection{Single specie ZRP}
The model is
defined as follows. Particles hop on a lattice with a rate $u(n_i(t))$
depending only on the number of particles at the considered site $i$
at the moment of the jump ;
$u$ is a function which vanishes at $0$ (no motion from site $i$ if no
particle). Here the continuum limit is obtained in a way slightly
different from before. The lattice spacing is $a$, and is sent to
zero, while the density is defined as before from the particle number.
It is now a straightforward exercice to generalize what has
been done previously to get the following action:
\begin{equation}
S[\{\rho,\rhoh\}]=-\int\rmd t\int\rmd^D x\left\{
\rhoh_x\left(-\partial_t \rho_x+\Delta u_a(\rho_x)\right)
+u_a(\rho_x)\left(\nabla\rhoh_x\right)^2\right\},
\end{equation}
with $u_a(\rho)=a^{-D}u(a^D\rho)$, 
after rescaling of the microscopic time $t\rightarrow
t/a^2$. Integrating out the $\rhoh$ field 
like previously, one gets that the dynamics of the density field
follows the SDE
\begin{eqnarray}\label{eqzrp}
\partial_t\rho_x&=&-\nabla\cdot J_x\\
J_x&=&-\nabla u_a(\rho_x)+\sqrt{u_a(\rho_x)}\xi_x,
\end{eqnarray}
where $\xi$ has the same statistics as in (\ref{eqn:dea}).
The steady state free energy can be obtained from the Fokker-Planck
equation. We introduce the instantaneous probability distribution of
paths:
\begin{equation}
{\cal P}[\{\rhot\},t]=\left\langle \prod_x \delta\left(
\rhot_x-\rho_x(t)\right)\right\rangle,
\end{equation}
which obeys the following Fokker-Planck equation:
\begin{equation}\label{eqn:fok}
\partial_t{\cal P}[\{\rhot\},t]=-\left\langle{\frac{\delta}{\delta\rhot_x},
\Delta u_a(\rhot_x){\cal P}[\{\rhot\},t]+
\nabla\cdot\left(u_a(\rhot_x)\nabla {\cal P}[\{\rhot\},t]\right)
}\right\rangle,
\end{equation}
Where we used the notation $\langle f,g\rangle=\int\rmd^D x\, f_x g_x$.
Writing formally
\begin{equation}\label{eqn:f}
\Delta u_a(\rhot_x)=\nabla\cdot\left(u_a(\rhot_x)\nabla \frac{\delta{\cal
    F}[\{\rhot\}]}{\delta\rhot_x}\right),
\end{equation}
Eq. (\ref{eqn:fok}) reads
\begin{equation}
\partial_t{\cal P}=-\left\langle{\frac{\delta}{\delta\rhot_x},
\nabla\cdot\left(u_a(\rhot_x)\nabla \frac{\delta{\cal
    F}[\{\rhot\}]}{\delta\rhot_x}\right){\cal P}+
\nabla\cdot\left(u_a(\rhot_x)\nabla {\cal P}\right)
}\right\rangle.
\end{equation}
It is straightforward to check that the steady state distribution is
obtained from the solution ${\cal F}$ of (\ref{eqn:f}), which is
\begin{equation}
{\cal F}[\{\rhot\}]=\int\rmd^D x\int_{1/a^D}^{\rhot_x}\rmd r \ln u_a(r),
\end{equation}
via
\be
P[\{\rhot\}]\propto \exp\left(-{\cal F[\{\rhot\}]}\right)
\ee
This is the continuous analog of
\begin{equation}
\ln\left(\prod_i u!_{n_i}\right),
\end{equation}
where $u!_n=\prod_{k=1}^n u_k$. The possibility to solve
(\ref{eqn:f}) by direct integration is related to the fact that
the steady state measure is factorized. In general, this measure
is not factorized, and it is more difficult to read ${\cal F}$ in
the Fokker-Planck equation. We will illustrate such difficulty in
the context of the two-species ZRP.

We want to end this section with a remark on the continuum stochastic equations
derived for the ZRP. They are valid on scales $l$ such that $a<<l$. On
the other hand there is no assumption on the correlation length $\xi$.
Coarse-graining the previous stochastic equations on scales much larger than
$\xi$ would lead to hydrodynamic equations (see the next sections).
The main interest of
eqs. (\ref{eqzrp}) is that they provide at the same time a continuum
description and can describe critical and out of equilibrium properties
of ZRP, ie. they have the same status as the model A stochastic equations
for the (non conserved) dynamics of the Ising model.

Remark that we have taken spatial derivatives of the fields
without justification of their smoothness. In order to make this
more rigorous, one would have to consider the density fields as
distributions acting on the space of smooth functions with compact
support on the discrete lattice. This is in fact a rigorous but
much less clear way of doing the coarse-graining which we refer
to.

\subsubsection{Two species ZRP}
This version of the ZRP involves two species $A$ and $B$, with
occupation numbers $n^A_i(t)$ and $n^B_i(t)$. The difference with
the single-species
case, is that the hopping rate from site $i$ of particles of a given
species is a function of the local number of particles of the other
species, which we denote by $u^A(n^A_i)$ and $u^B(n^B_i)$.
In the continuum limit, the density fields verify the SDE
\begin{eqnarray}
\partial_t\rho^A_x&=&
\Delta u^A_a(\rho^A_x,\rho^B_x)+\nabla\cdot\left(\sqrt{u^A_a(\rho^A_x,\rho^B_x)}\xi^A_x\right),\\
\partial_t\rho^B_x&=&
\Delta u^B_a(\rho^A_x,\rho^B)+\nabla\cdot\left(\sqrt{u^B_a(\rho^A_x,\rho^B)}\xi^B_x\right).
\end{eqnarray}
If the steady state is factorized, the following conditions must be
verified:
\begin{eqnarray}
\ln u^A_a(\rhot^A_x,\rhot^B_x)&=&\frac{\delta{\cal
    F}}{\delta\rhot^A_x}+{\rm constant},\\
\ln u^B_a(\rhot^A_x,\rhot^B_x)&=&\frac{\delta{\cal
    F}}{\delta\rhot^B_x}+{\rm constant}. 
\end{eqnarray}
This gives a necessary condition for factorization of the steady
state:
\begin{equation}
u^B_a(\rhot^A_x,\rhot^B_x)\frac{\partial u^A_a(\rhot^A_x,\rhot^B_x)}{\partial\rhot^B_x}=
u^A_a(\rhot^A_x,\rhot^B_x)\frac{\partial u^B_a(\rhot^A_x,\rhot^B_x)}{\partial\rhot^A_x}.
\end{equation}
This is the continuous analog of the condition given by Evans and Hanney for
the model on-lattice~\cite{evab}.

\subsection{Hydrodynamic limits}
We now show how hydrodynamic limits can be obtained within our
field theory approach. Taking hydrodynamic limits consists in
splitting the volume of the system into boxes equal sizes, with
volumes a fraction of the total volume of the system. This amounts
to using a coarse-graining length $l$ much smaller than the system
size $L$, but much larger than the lattice spacing $a$. In one
dimension, boxes are thus located between $xL$ and $xL+l$, with
$x\in\{0,l/L,2l/L,\cdots,(L-l)/L\}$. In the hydrodynamic limit,
where both $l$ and $L$ go to infinity, $x$ becomes a real
coordinate in $[0,1]$ (this corresponds to $dx=l/L$). In order to
avoid artificial showness, the time is rescaled: $t\rightarrow
t/L^2$. The hydrodynamic limit is in principle different from the
much less controlled continuum limit taken before. There are in
fact two possibilities. The first one is that the hydrodynamic limit exists, in
which case the limits are connected ; this happens for instance
away from criticality, where correlation lengths are finite. In
this case the large time and length scale behavior of the
continuum field theory leads to the hydrodynamic limit. The other
possibility is that the hydrodynamic limit does not exist, while
the continuum limit of the field theory still exists, although
great care must be taken; this typically happens at criticality
where the field theory in the continuum allows for renormalization
group calculations.

There are now standard routes for rigorously deriving hydrodynamic
limits for particle systems on lattice, such as ZRP or exclusion
processes. We will not repeat them, as they can be found in
textbooks and recent papers. Instead, we will show how our field
theoretic approach naturally leads to hydrodynamic equations and
large deviation functionals. Although the following results are
not new, we think that the derivation we present gives an
interesting new perspective on these problems.

The basic idea underlying hydrodynamic equations is the existence
of a law of large numbers for density profiles. Let us imagine a
system with a finite correlation length $\xi$, coarse-grained
using a coarse-graining length $l\gg\xi$. Inside all boxes, the
density field is the sum of a large number of independent
identically distributed processes and is thus Gaussian. It can
thus be characterized by first and second moments, and a large
deviation functional is easily deriveed in terms of the macroscopic 
transport coefficients - see last section.

\subsubsection{Zero Range and Exclusion Processes}
Let's start by a simple example, single species ZRP in one
dimension. In such models, the stationary measure is factorized,
which makes it possible and simple to compute local averages from
the marginals of the total joined distribution. Coarse-graining on
the scale $l$ is equivalent to averaging out all fields while
fixing the value of the density inside each boxe at each time to
its average value $\rhob(x,t)$. This amounts to using constant
conjugated field $\nh$ inside each box. However, instead of
restricting the path integral to piecewise constant $\nh$ fields,
it is simpler to restrict it to slow varying $\nh$. More
precisely, we restrict the measure to fields which 
significant variations occur on scales larger than $l$. The dynamical
action reads when the lattice spacing $a=1/L$ goes to zero:
\begin{equation}
  -\sum_i\int\rmd t\,\left\{
-\nh_i\partial_t n_i+a^2 u(n_i)\left[
(\nabla\nh_i)^2+\Delta\nh_i
\right]
\right\}
\end{equation}
Now comes the fundamental hypothesis of local equilibrium. The
variation of the total number of particle inside large boxes is
only due to particle flows at its boundaries. If the size is large
enough, one may assume that the time scale for any significant
variation is also large and the system is locally at equilibrium.
The local average density evolves slowly following the
hydrodynamic equation which we want to derive, and we make the
following approximation:
\begin{equation}
\fl\frac{1}{l}\sum_{i\in B_x}u(n_i(t)) \left[
(\nabla\nh_i(t))^2+\Delta\nh_i(t)
\right]\approx  \frac{1}{l}\sum_{i\in B_x}\overline{u(n)}_x(t) \left[
(\nabla\nh_i(t))^2+\Delta\nh_i(t)
\right],
\end{equation}
where
\begin{equation}
{\overline{u(n)}}_x(t)=\frac{1}{l}\sum_{i\in B_x} u(n_i(t))
\end{equation}
is the local mean value of $u(n_i(t))$. The assumption of local
equilibrium also gives this mean value:
\begin{equation}
\overline{u(n)}_x(t)=\langle u(n)\rangle_{\rhob_x(t)},
\end{equation}
where $\langle\cdot\rangle_{\rhob}$ stands for the average using the
marginal with mean $\rhob$.

It can be obtained from the grand canonical partition function
\begin{equation}
  \Theta(z)=\sum_n p_n z^n,
\end{equation}
with $p_n=\frac{1}{u!_n}$:
\begin{equation}\label{eqn:gc}
  \rhob=z\frac{\partial \Theta}{\partial z}.
\end{equation}
Inverting (\ref{eqn:gc}) defines $R(\rhob)=z$. Then we get:
\begin{equation}
  \overline{u(n)}_{\rhob}=z=R(\rhob).
\end{equation}
In addition, $\nh$ being a very slow varying field, one has in one
dimension:
\begin{equation}
  \sum_{i\in B_x}\left[(\nabla\nh_i)^2+\Delta\nh_i\right]\approx
  l\left[\left(\nabla_x\rhoh_x\right)^2+\Delta\rhoh_x \right],
\end{equation}
where $\nh_i\approx \rhoh_x$ inside the box of size $l$ around $x$ and
$\nabla_x$ is now the gradient at the intermediate scale $x$, i.e.
the variation between two points at distance $l$ divided by $l$.
Carrying the same analysis with the term $-\sum_i\nh_i\partial_t n_i$,
and rescaling the time by $a^2=1/L^2$ the action finally reads:
\begin{equation}\label{eqn:hyd}
S_{\rm hydro}=-L\int\rmd t\int_0^1\rmd x\,\left\{
\rhoh_x\partial_t \rhob_x+\rhoh_x \Delta
R(\rhob_x)+R(\rhob_x)\left(\nabla\rhoh_x\right)^2
\right\}.
\end{equation}
The result (\ref{eqn:hyd}) is a good illustration of the emergence of
a law of large numbers at the hydrodynamic level, where the
empirical current has mean $\nabla R(\rhob)$ and mean square $R(\rhob)$.
This can be seen either from a direct interpretation in terms of an
effective Langevin equation or by using (\ref{eqn:gen}) and second
derivatives of the generating function. In particular (\ref{eqn:gen})
gives the hydrodynamic equation:
\begin{equation}
  \partial_t\rhob=\Delta R(\rhob).
\end{equation}
The action (\ref{eqn:hyd}) gives the large deviation functional
for density fluctuations, by integration on the field $\rhoh$. If
one formally defines
$\sigma(\rhob)=-\nabla\cdot\left(R(\rhob)\nabla\right)$, then the
weight of any - coarse-grained - trajectory is given by
\begin{equation}
  P[\{\rhob\}]\propto\int\,\{d\rhoh_x(t)\}e^{-S_{\rm hydro}[\{\rhob,\rhoh\}]},
\end{equation}
which gives
\begin{equation}
  P[\{\rhob\}]\propto e^{-{\tilde S}_{\rm hydro}[\{\rhob\}]},
\end{equation}
with
\begin{equation}
  {\tilde S}_{\rm hydro}[\{\rhob\}]=L\int\rmd t\int_0^1\langle\partial_t
    \rhob_x-\Delta R(\rhob_x),\sigma(\rhob_x)^{-1}\left(\partial_t
    \rhob_x-\Delta R(\rhob_x)\right)\rangle.
\end{equation}
Due to the $L$ factor in front of the integral, the path integral
is dominated by the hydrodynamic trajectory, which minimizes the
large deviation functional.

The case of ZRP is relatively simple as the stationary
distribution is factorized, and thus $\overline{u(n)}$ is easily
computed. In general, the whole product measure must be used.

Starting from (\ref{eqn:sep}), the fluctuating hydrodynamic equation
is obtained
for exclusion processes, where a drift velocity $v$ can be added:
\begin{equation}
  \partial_t\rhob=-\nabla\cdot\left(v\rhob(1-\rhob)\right)+\Delta\rhob
+\nabla\cdot\left(\sqrt{2\rhob(1-\rhob)/L}\,\eta\right),
\end{equation}
where $\eta$ is as usual a normal Gaussian noise field. This
equations contains initial symmetries of the microscopic model,
i.e. particle-hole symmetry.

\subsection{Systems with reaction processes}
Reaction processes can provide an
example of hydrodynamic limits
in which the time has not to be rescaled because there are
no conserved quantities.

Again, we take for canonical example the pair annihilation
$A+A\rightarrow\emptyset$, but without diffusion. Using the notations
used until here, one gets
\begin{eqnarray}
\lambda\sum_i\int\rmd s\,n_i(s)(n_i(s)-1)\left(e^{-2\nh_i(s)}-1\right)&\approx
&L\lambda\int\rmd t\int_0^1\rmd^D x\langle
n(n-1)\rangle_{\rhob_x(s)}\nonumber\\
&\times&\left(e^{-2\rhoh_x(s)}-1\right)
\end{eqnarray}
and thus \be\label{eqn:hre}\fl S_{\rm dyn}\approx -L\int\rmd
s\int_0^1\rmd^D x\left\{ \lambda\langle n(n-1)\rangle_{\rhob_x(s)}
\left(e^{-2\rhoh_x(s)}-1\right) -\rhoh_x(s)\partial_s\rhob_x(s)
\right\}. \ee 
The difficulty is still to compute the static
average. For the simple example given here, there is no steady
state except the empty one, but occupation numbers remain
Poissonian at all times, and thus \be \langle
n(n-1)\rangle_{\rhob(s)}=\rhob(s)^2. \ee As $L$ is large, the
measure is dominated by saddle points of the action. Thus the
hydrodynamic equation is
\begin{equation}
\partial_t\rhob_x(t)=-2\lambda\rhob_x(t)^2 e^{-2\rhoh_x(t)},
\end{equation}
where $\rhoh_x(t)$ is solution of
\begin{equation}
\partial_t\rhoh_x(t)=-2\lambda\rhob_x(t)\left(e^{-2\rhoh_x(t)}-1\right).
\end{equation}
This equation describes the behavior of the system on large length
scales but timescales of order of one. There is no need of rescaling the time
because the density is not conserved so the coarse-grained density
still evolves on timescales of order one.

\subsection{Large Deviation Functional}
Recently there have been a lot of interest and works on large
deviation functional for out of equilibrium driven systems. In the
following we would like to show how this large deviation
functional appears naturally within our framework. Our approach is
not rigorous compared to the previous ones \cite{ber,der}. On the
other hand it shows clearly in our opinion the key ingredients and
it can be easily generalized to systems more complicated than the
ones considered up to now. In the following, for the sake of
simplicity, we will focus on one dimensional driven stochastic
lattice gases. The driving can be due either to the boundary
conditions or to a force not deriving from a potential.

We shall focus on the probability that the system follows a given
path $\{n(x,t)\}$ in configuration space. This can be obtained
formally by integrating out the auxiliary field $\hat n$ but this is in
principle not feasible. However, if one is only interested in the
probability of a given path after coarse-graining then the tasks
simplifies a lot. Let us rescale time and length scales in the
hydrodynamic way described before and focus on the probability of
smooth paths. This can be obtained with a construction very
similar to the one used to obtain Feynman path integral in quantum
mechanics:
\be
P[\{\rho(x,t)\}]\simeq \prod_{\Delta x, \Delta t} P_{\Delta x}[\rho; t\rightarrow t+\Delta t]
\ee
where $P_{\Delta x}[\rho; t\rightarrow t+\Delta t]$ is the
probability inside a small coarse-grained box of the evolution of
the density on a small coarse-grained time ($\rho$ is the notation
for the coarse-grained density). If the hydrodynamic limit exists
then each $P_{\Delta x}[\rho; t\rightarrow t+\Delta t]$ can be
replaced by its hydrodynamic expression because although the
global fluctuation of the density can be large, on each small
(coarse-grained) box they are small. This expression equals the
corresponding current distribution $P_{\Delta x}[J; t\rightarrow
  t+\Delta t]$, up to a constant Jacobian. Repeating the procedure 
explained previously for the hydrodynamic limit one obtains a
field theory such that the auxiliary field can be integrated out
leading to:
\be
\fl P_{\Delta x} [J; t\rightarrow t+\Delta t]\simeq\int {\cal D} \rho
\exp\left[ -L \Delta x \Delta t \left(J-J_{\rm
    av}(\rho)\right)^2/\chi(\rho)\right]
\ee
where $J_{\rm av}(\rho)=-D(\rho)\nabla \rho+\chi(\rho)E$ is the
empirical current; $D,\chi$ are the
transport coefficients characterizing the hydrodynamic limit,
i.e. diffusion coefficient and mobility ($L$ is the size of the
system) and $E$ is the external driving field. 
Putting together all the probability on small boxes one obtains the
large deviation functional: 
\be\fl
P[\{\rho(x,t)\}]\simeq \int {\cal D} \rho
\exp \left[-L \int dt \langle \partial_t \rho+\nabla\cdot
  J_{\rm av}(\rho),\eta(\rho)\left(\partial_t \rho+\nabla\cdot J_{\rm av}(\rho)\right)\rangle\right],
\ee
where $\eta(\rho)$ is the inverse of the operator
$-\nabla\cdot(\chi(\rho)\nabla)$. 
Note that this field theory corresponds to a stochastic equations for the density field:
\be
\partial_t \rho(x,t)=-\nabla\cdot J_{\rm av}(\rho(x,t))+\nabla\cdot
	[\sqrt{\chi(\rho(x,t))}\, \eta(x,t)]
\ee
where $\eta$ is a white noise in space and time,
$\langle\eta(x,t)\eta(x',t')\rangle=2\delta(t-t')\delta(x-x')/L$, 
and the multiplicative noise must be interpreted in the Ito sense.

This derivation although certainly not rigorous has the virtue of
showing in a simple manner why only hydrodynamic transport
coefficients matter for the derivation of the large deviation
functional and why this is related to a stochastic equation which
leads to a non-linear fluctuating hydrodynamics. It can be
straightforwardly generalized to systems with a  more complicated
hydrodynamics, e.g. real fluids where density, energy and momentum
are conserved.

\section{Conclusion}
The aim of
 this paper was to discuss and present
a field theoretical approach to interacting particle systems.
We wanted to show a dual version of the Doi-Peliti field theory that
can be obtained from the stochastic process and that is directly
 related to stochastic equations. The advantage of this field theory
is that it focuses on the physical density fields as we have shown in
 some applications, e.g. ZRP continuum stochastic equations, large
 deviation functional, 
hydrodynamic limits. We think that this approach will help to tackle
difficult problems in interacting particle systems.

\section*{Acknowledgements} 
This paper is based on a joint work with
A. Andreanov and J.-P. Bouchaud. We thanks C. Godreche for
discussion especially on the ZRP process. We are grateful to the
organizers of the semester "Principles of the dynamics of
Nonequilibrium systems" hold at the Isaac Newton Institute for
Mathematical Sciences for giving us the opportunity of presenting
our work.

\end{document}